# EFFECT OF BOTH Z AND $Z'$-MEDIATED FLAVOR-CHANGING NEUTRAL CURERENTS ON $B_s \to \tau^+\tau^-$ DECAY

S. Sahoo[1*], D. Banerjee[1], M. Kumar[1] and S. Mohanty[2]

[1]Department of Physics, National Institute of Technology,
Durgapur-713209, West Bengal, India
*E-mail: sukadevsahoo@yahoo.com

[2]Department of Physics, Trident Academy of Technology,
Bhubaneswar – 751024, Orissa, India.

**Abstract**

In recent years, $B_s \to \tau^+\tau^-$ rare decay has attracted a lot of attention since it is very sensitive to the structure of standard model (SM) and potential source of new physics beyond SM. In this paper, we study the effect of both Z and $Z'$-mediated flavor-changing neutral currents on the $B_s \to \tau^+\tau^-$ decay. We find the branching ratio $B(B_s \to \tau^+\tau^-)$ is enhanced relative to SM prediction, which would help to explain the recent observed CP-violation from like-sign dimuon charge asymmetry in the B system.

## 1. Introduction

In recent years B-physics has been a very active area of research both experimentally and theoretically [1] because it is one of the domains where new physics (NP) might very well reveal itself. Rare B decays [2–4] induced by flavor-changing neutral current (FCNC) transitions are very important to probe the flavor sector of the SM. In the SM they arise from one-loop diagrams and are generally suppressed in comparison with the tree diagrams. Nevertheless, one-loop FCNC processes can be enhanced by orders of magnitude in some cases due to the presence of new physics. New physics comes into action in rare B decays in two different ways : either (a) through a new contribution to the Wilson coefficients or (b) through a new structure in the effective Hamiltonian, which are both absent in the SM. Rare B decays which are mediated by FCNC transitions are of two kinds: (i) $b \to s$, and (ii) $b \to d$. The $B_s \to \ell^+\ell^-$ ($\ell = \mu, \tau$) decay involves $b \to s$ transitions. The SM produces only the combinations $\ell_R^+ \ell_L^-$ and $\ell_L^+ \ell_R^-$. These decays are highly suppressed in the SM. However, they can be significantly enhanced in many scenarios beyond the SM [2,5–9]. Since $B_s \to \tau^+\tau^-$ decay has not been observed, one can only constrain model parameters based on the measured limits of branching ratio.

Furthermore, the recent observation of a sizable CP-violation from like-sign dimuon charge asymmetry in the B system [10–12] by the D0 collaboration has attracted tremendous theoretical interest. The updated observed value [12]:

$$A_{s\ell}^b = (-7.87 \pm 1.72 \pm 0.93) \times 10^{-3} \qquad (1)$$

is significantly larger than the SM prediction [13–15]

$$\left(A_{s\ell}^b\right)_{SM} = (-2.3 \pm 0.4) \times 10^{-4} \qquad (2)$$

Such an asymmetry can be used as a probe of flavor structure of new physics [16]. The dimuon charge asymmetry can be written as a linear combination of the semileptonic asymmetries $a_{s\ell}^d$ and $a_{s\ell}^s$ in $B_d$ and $B_s$ decays respectively [12–15]:

$$A_{s\ell}^b = (0.594 \pm 0.022) a_{s\ell}^d + (0.406 \pm 0.022) a_{s\ell}^s \qquad (3)$$

These asymmetries $a_{s\ell}^q$ (q = d,s) are related to the mass and width differences in the $B_q - \overline{B_q}$ system as [17,18]:

$$a_{s\ell}^q = \mathrm{Im}\frac{\Gamma_{12}^q}{M_{12}^q} = \frac{|\Gamma_{12}^q|}{|M_{12}^q|}\sin\phi_q = \frac{\Delta\Gamma_q}{\Delta M_q}\tan\phi_q. \qquad (4)$$

In the SM, since $\tan\phi_d \approx 0.075$, a large enhancement requires fine tunning in $B_d$ sector. Again, an enhancement in $\Delta\Gamma_d$ would imply the NP contribution to the branching ratio of $B_d$ decay modes to a few percent, which is ruled out by the experimental results [19]. On the other hand, in the SM, $\phi_s \approx 0.004$ [13–15], and the branching ratios of some decay modes, such as $B_s \to \tau^+\tau^-$, have not yet been strongly constrained. New physics models that increase the decay rate of $B_s \to \tau^+\tau^-$ contribute to the absorptive part of $B_s - \overline{B}_s$ mixing and may enhance $\Delta\Gamma_s$ [20,21]. The enhancement in $\Delta\Gamma_s$ corresponds to an enhancement in the branching ratio $B(B_s \to \tau^+\tau^-)$. Therefore, a measurement of $B(B_s \to \tau^+\tau^-)$ gives a better understanding of new physics involved in $B_s - \overline{B}_s$ mixing. The study of $B_s \to \tau^+\tau^-$ decay [20,21] would explain the observed like-sign dimuon charge asymmetry in the B system. In this paper, we study $B_s \to \tau^+\tau^-$ decay considering the effect of both Z and $Z'$-mediated FCNCs that change the effective Hamiltonian and enhance the branching ratio.

$Z'$ bosons are known to exist naturally in well-motivated extensions of the SM [22–27]. Theoretically it is predicted that they exist in grand unified theories (GUTs), left-right symmetric models, Little Higgs models, superstring theories and theories with large extra dimensions. But experimentally $Z'$ boson is not conclusively discovered so far. Hence, the exact mass of $Z'$ boson is not known. However, there are stringent limits on the mass of an extra $Z'$ from the non-observation of direct production followed by decays into $e^+e^-$ or $\mu^+\mu^-$ by CDF [28,29], while indirect constraints from the precision data also limit the $Z'$ mass (weak neutral current processes and LEP II) and severely constrain the $Z-Z'$ mixing angle $\theta$ [30–32]. These limits are model-dependent, but are typically in the range $M_{Z'} \geq 500\ GeV$ and $|\theta| \leq 10^{-3}$ for standard GUT models; stringent strong constraints on $M_{Z'}$, of the order of 1 TeV, are obtained in models with nonuniversal flavor gauge interactions [33–36].

In the $Z'$ sector, there has been a great deal of investigation to understand the underlying physics beyond the SM [37–43]. It has been shown that a leptophobic $Z'$ boson can appear in $E_6$ gauge models due to mixing of gauge kinetic terms [44–48]. Flavor mixing can be induced at the tree level in the up-type and/or down-type quark sector after diagonalizing their mass matrices. Mixing between ordinary and exotic left-handed quarks induces Z-mediated FCNCs. The right-handed quarks $d_R, s_R$ and $b_R$ have different $U(1)'$ quantum numbers than exotic $q_R$ and their mixing will induce $Z'$-mediated FCNCs [44–52] among the ordinary down quark types. Tree level FCNC interactions can also be induced by an additional $Z'$ boson on the up-type quark sector [53]. In the $Z'$ model [54], the FCNC $b-s-Z'$ coupling is related to the flavor-diagonal couplings $qqZ'$ in a predictive way, which is then used to obtain upper limits on the leptonic $\ell\ell Z'$ couplings. Hence, it is possible to predict the branching ratio for the taunic decay of the $B_s$. With FCNCs, both Z and $Z'$ boson contribute at tree level, and their contribution will interfere with the SM contributions.

The rest of the paper is structured as follows: in Section 2, we discuss the $B_s \to \tau^+\tau^-$ decay in the SM. In Section 3, we give a brief introduction of the model. Then, we evaluate the effective Hamiltonian for $B_s \to \tau^+\tau^-$ decay considering the Feynman diagram and the contribution from both the Z and $Z'$ bosons. In Section 4, we calculate the branching ratio for $B_s \to \tau^+\tau^-$ decay. Then we discuss the results and compare with others.

## 2. $B_S \to \tau^+\tau^-$ Decay in the Standard Model

In the SM, the $B_s \to \tau^+\tau^-$ process is loop-suppressed. However, it is potentially sensitive to new physics beyond the SM. This decay involves $b \to s$ transitions. The effective Hamiltonian [2] describing the process $B_s \to \ell^+\ell^-\ (\ell = \mu, \tau)$, is

$$H_{eff} = \frac{G_F \alpha}{\sqrt{2}\pi} \lambda_t \left[ C_9^{eff} \left( \bar{s} \gamma^\mu P_L b \right) \left( \bar{\ell} \gamma^\mu \ell \right) + C_{10} \left( \bar{s} \gamma^\mu P_L b \right) \left( \bar{\ell} \gamma^\mu \gamma_5 \ell \right) \right.$$
$$\left. - \frac{2 C_7 m_b}{p^2} \left( \bar{s} \not{p} \gamma^\mu P_R b \right) \left( \bar{\ell} \gamma^\mu \gamma_5 \ell \right) \right], \tag{5}$$

where $G_F$ is the Fermi coupling constant, $\lambda_t = V_{tb} V_{ts}^*$, $P_{R,L} = \frac{1}{2}(1 \pm \gamma_5)$, $p = p_+ + p_-$ the sum of the momenta of the $\ell^+$ and $\ell^-$, and $C_7$, $C_9^{eff}$ and $C_{10}$ are Wilson coefficients [9, 55–58] evaluated at the b quark mass scale.

We use the Vacuum Insertion Method (VIM) [59] for the evaluation of matrix elements as:

$$\left\langle 0 \left| \bar{s} \gamma^\mu \gamma_5 b \right| B_s^0 \right\rangle = i f_{B_s} p_B^\mu, \tag{6}$$

$$\left\langle 0 \left| \bar{s} \gamma_5 b \right| B_s^0 \right\rangle = i f_{B_s} m_{B_s}, \tag{7}$$

and

$$\left\langle 0 \left| \bar{s} \sigma^{\mu\nu} P_R b \right| B_s^0 \right\rangle = 0. \tag{8}$$

Let us consider the contribution of each term in equation (5). $p_B^\mu = p_+^\mu + p_-^\mu$, hence the contribution from $C_9$ term in equation (5) will vanish upon contraction with the lepton bilinear, $C_7$ will also give zero by equation (8) and the remaining $C_{10}$ term will get a factor of $2 m_\ell$. Using the above results, we can write the transition amplitude for this process as

$$M\left(B_s \to \ell^+ \ell^-\right) = i \frac{G_F \alpha}{\sqrt{2}\pi} \lambda_t f_{B_s} C_{10} m_\ell \left( \bar{\ell} \gamma_5 \ell \right), \tag{9}$$

and the corresponding branching ratio [2, 9, 55–58] is given by

$$B\left(B_s \to \ell^+ \ell^-\right) = \frac{G_F^2 \tau_{B_s}}{16 \pi^3} \alpha^2 f_{B_s}^2 m_{B_s} m_\ell^2 \left| V_{tb} V_{ts}^* \right|^2 C_{10}^2 \sqrt{1 - \frac{4 m_\ell^2}{m_{B_s}^2}}. \tag{10}$$

The value of the branching ratio in the standard model is predicted as

$$B\left(B_s \to \tau^+ \tau^-\right) = 7.7 - 8.0 \times 10^{-7} \; [1,7,8,21]. \tag{11}$$

The experimental study of $B_s \to \tau^+ \tau^-$ decay is quite challenging due to the difficulty in identifying $\tau$'s. This decay has not been fully observed. However, the LHCb collaboration [60] has allowed this branching ratio up to 3.5 % without considering the NP contribution to $B_d$ decays, i.e.

$$B\left(B_s \to \tau^+ \tau^-\right) \leq 3.5 \times 10^{-2} \tag{12}$$

If NP contribution to $B_d$ decays is considered, this bound may be further changed.

## 3. The Model

In extended quark sector model [61–64], besides the three standard generations of the quarks, there is an $SU(2)_L$ singlet of charge $-1/3$. This model allows for Z-mediated FCNCs. The up quark sector interaction eigenstates are identified with mass eigenstates but down quark sector interaction eigenstates are related to the mass eigenstates by a 4 × 4 unitary matrix, which is denoted by K. The charged-current interactions are described by

$$L^W_{int} = \frac{g}{\sqrt{2}} \left( W^-_\mu J^{\mu^+} + W^+_\mu J^{\mu^-} \right), \tag{13}$$

$$J^{\mu^-} = V_{ij} \bar{u}_{iL} \gamma^\mu d_{jL}. \tag{14}$$

The charged-current mixing matrix V is a 3 × 4 submatrix of K :

$$V_{ij} = K_{ij} \quad \text{for } i = 1,......3, \quad j = 1,......,..4. \tag{15}$$

Here, V is parametrized by six real angles and three phases, instead of three angles and one phase in the original CKM matrix.

The neutral-current interactions are described by

$$L^Z_{int} = \frac{g}{\cos\theta_W} Z_\mu \left( J^{\mu 3} - \sin^2\theta_W J^\mu_{em} \right), \tag{16}$$

$$J^{\mu 3} = -\frac{1}{2} U_{pq} \bar{d}_{pL} \gamma^\mu d_{qL} + \frac{1}{2} \delta_{ij} \bar{u}_{iL} \gamma^\mu u_{jL}. \tag{17}$$

In neutral-current mixing, the matrix for the down sector is $U = V^\dagger V$. Since V is not unitary, $U \neq 1$, the nondiagonal elements do not vanish:

$$U_{pq} = - K^*_{4p} K_{4q} \quad \text{for } p \neq q. \tag{18}$$

The various $U_{pq}$ are non-vanishing, which allow for flavor-changing neutral currents that would be a signal for new physics.

Now considering the $B_s \to \tau^+\tau^-$ decay in the presence of Z-mediated FCNC [61–63] at tree level (Fig.1). The $Zbs$ FCNC coupling, which affects B-decays, is parameterized by one independent parameter $U_{sb}$ and this parameter is constrained by branching ratio of the decay $B_s \to \tau^+\tau^-$. Given that the Z boson contributes to $B_s \to \ell^+\ell^-$ ($\ell = \mu, \tau$), one can write the effective Hamiltonian [2] as:

$$H_{eff}(Z) = \frac{G_F}{\sqrt{2}} U_{sb} \left[ \bar{s}\gamma^\mu(1-\gamma_5)b \right]\left[ \bar{\ell}\left( C^\ell_V \gamma_\mu - C^\ell_A \gamma_\mu \gamma_5 \right)\ell \right], \tag{19}$$

where $C^\ell_V$ and $C^\ell_A$ are the vector and axial vector $Z\ell^+\ell^-$ couplings and are given as

$$C_V^\ell = -\frac{1}{2} + 2\sin^2\theta_W, \quad C_A^\ell = -\frac{1}{2}. \tag{20}$$

The transition amplitude is given as

$$M(B_s \to \ell^+\ell^-) = -i\frac{G_F}{\sqrt{2}} U_{sb} f_{B_s} C_A^\ell 2 m_\ell (\bar{\ell}\gamma_5 \ell), \tag{21}$$

and the corresponding branching ratio is given as

$$B(B_s \to \ell^+\ell^-)_Z = \frac{G_F^2 \tau_{B_s}}{4\pi} |U_{sb}|^2 f_{B_s}^2 m_{B_s} m_\ell^2 |C_A^\ell|^2 \sqrt{1 - \frac{4m_\ell^2}{m_{B_s}^2}}. \tag{22}$$

The same idea can be applied to a $Z'$ boson i.e., mixing among particles which have different $Z'$ quantum numbers will induce FCNCs due to $Z'$ exchange [49–52,65,66] and these effects can be as large as Z-mediated FCNCs. Since the $U_{pq}^Z$ are generated by mixing that breaks weak isospin, they are expected to be at most $O(M_1/M_2)$, where $M_1(M_2)$ is typical light (heavy) fermion mass. But the $Z'$-mediated coupling $U_{pq}^{Z'}$ can be generated via mixing of particles with same weak isospin and are not suppressed by the mass of heavy fermion. Even though $Z'$-mediated interactions are suppressed relative to Z, these are compensated by the factor $U_{pq}^{Z'}/U_{pq}^Z \sim (M_2/M_1)$. Thus the effect of $Z'$-mediated FCNCs are comparable to that of Z-mediated FCNCs. If we assume $|U_{sb}^{Z'}| \sim |V_{tb} V_{ts}^*|$, then it is possible to write $U_{sb}$ instead of $U_{sb}^{Z'}$, which gives significant contributions to the $B_s \to \tau^+\tau^-$ decay. The new contributions from $Z'$ boson have similar effect as from the Z boson. Therefore, we write the general effective Hamiltonian that contributes to $B_s \to \ell^+\ell^-$, in the light of equation (19) as :

$$H_{eff}(Z') = \frac{G_F}{\sqrt{2}} U_{sb} [\bar{s}\gamma^\mu(1-\gamma_5)b][\bar{\ell}(C_V^\ell \gamma_\mu - C_A^\ell \gamma_\mu \gamma_5)\ell] \left(\frac{g'}{g}\frac{M_Z}{M_{Z'}}\right)^2, \tag{23}$$

where $g = e/(\sin\theta_W \cos\theta_W)$ and $g'$ is the gauge coupling associated with the $U(1)'$ group. The net effective Hamiltonian can be written, from equation (19) and (23), as $H_{eff} = H_{eff}(Z) + H_{eff}(Z')$ and

$$H_{eff} = \frac{G_F}{\sqrt{2}} U_{sb} [\bar{s}\gamma^\mu(1-\gamma_5)b][\bar{\ell}(C_V^\ell \gamma_\mu - C_A^\ell \gamma_\mu \gamma_5)\ell] \left[1 + \left(\frac{g'}{g}\frac{M_Z}{M_{Z'}}\right)^2\right], \tag{24}$$

and the corresponding branching ratio is given as

$$B\left(B_s \to \ell^+\ell^-\right)_{Z+Z'} = \frac{G_F^2 \tau_{B_s}}{4\pi} \left|U_{sb}\right|^2 f_{B_s}^2 m_{B_s} m_\ell^2 \left|C_A^\ell\right|^2 \sqrt{1-\frac{4m_\ell^2}{m_{B_s}^2}} \left[1+\left(\frac{g'}{g}\frac{M_Z}{M_{Z'}}\right)^2\right]^2.$$

(25)

This formula can be used for the calculation of branching ratio for the rare decays $B_s \to \ell^+\ell^-$ ($\ell = \mu, \tau$). We have already studied the case for $\ell = \mu$ [67]. In this paper, we study for $\ell = \tau$ and calculate the branching ratio in the next section.

## 4. Numerical Results and Discussions

In this section, we calculate the branching ratio for $B_s \to \tau^+\tau^-$ decay using the recent data [68]: $m_\tau = 1776.82 \pm 0.16$ MeV, $m_{B_s} = 5.3667 \pm 0.0024$ GeV, average $B_s$ lifetime $\tau_{B_s} = (1.497 \pm 0.015) \times 10^{-12}$ s, decay constant $f_{B_s} = 0.24$ GeV, $M_Z = 91.1876$ GeV, $G_F = 1.16639 \times 10^{-5}$ GeV$^{-2}$, $\sin^2\theta_W = 0.23$ and $\left|U_{sb}\right| \cong 10^{-3}$ [69,70]. Since the $Z'$ has not yet been discovered, its mass is unknown. A broad class of supersymmetric extensions of the SM predicts a $Z'$ boson mass lies in the range 250 GeV $< M_{Z'} <$ 2 TeV [71]. However, the lower mass limit can be as low as 130 GeV [72] if the coupling is weak. In a study of B meson decays with $Z'$-mediated FCNCs, Bargar et al. [73] study the $Z'$-boson in the mass range of a few hundred GeV to 1 TeV. At Tevatron [74], a light $Z'$ boson with a mass of approximately 150 GeV could explain the anomalies like top-quark forward-backward asymmetry. From the recent LHC results [75] the mass limits for the sequential standard model $Z'$ are about 1.49–1.69 TeV (CMS) and 1.77–1.96 TeV (ATLAS). Our estimation of the mass of $Z'$ boson from $B_q - \overline{B_q}$ mixing lies in the range of 1352–1665 GeV [76].

In general, the value of $g'/g$ is undetermined [77]. However, one expects that $g'/g \approx 1$ if both U(1) groups have the same origin from some GUT. We take $g'/g \approx 1$ in our calculations. From Eq. (25), it is clear that lower is the mass of $Z'$ boson, higher is the branching ratio. For an enhancement in the branching ratio $B(B_s \to \tau^+\tau^-)$, there is an enhancement in $\Delta\Gamma_s$. From Eq. (4), it is clear that the study of $B_s \to \tau^+\tau^-$ decay [20,21] would explain the observed like-sign dimuon charge asymmetry in the B system. In this paper, we take the mass of $Z'$ boson in the range of 100 GeV – 2 TeV for our calculations. For $M_{Z'} = 100$ GeV, we get

$$B\left(B_s \to \tau^+\tau^-\right)_{Z+Z'} = 1.64 \times 10^{-4}.$$

(26)

and for $M_{Z'} = 2$ TeV, we get

$$B\left(B_s \to \tau^+\tau^-\right)_{Z+Z'} = 9.0 \times 10^{-5}$$

(27)

From equation (26) and (27), it is clear that depending on the precise value of $M_{Z'}$, the $Z'$-mediated FCNCs gives sizable contributions to $B_s \to \tau^+\tau^-$ decay process. As the mass of $Z'$ boson increases, the contribution of $Z'$ boson to the branching ratio decreases. For other intermediate values of $M_{Z'}$, one can get the corresponding branching ratios. Mohanta [2] has already predicted the value of branching ratio as:

$$B(B_s \to \tau^+\tau^-)_Z = 8.9 \times 10^{-6}. \tag{28}$$

Our result is higher because the author in Ref. [2] considered only the effect of Z boson whereas we have considered the effect of both Z and $Z'$-mediated FCNCs on $B_s \to \tau^+\tau^-$ decay. From equations (26) and (27), it is clear that our estimated branching ratio for $B_s \to \tau^+\tau^-$ decay is enhanced relative to SM prediction [equation (11)]. Therefore, the $B_s \to \tau^+\tau^-$ decay process could provide signals for new physics beyond the SM. Our predicted branching ratio values match the recent LHCb collaboration [60] bound given in Eq. (12). This branching ratio can reach up to $10^{-4}$ in a two-Higgs doublet model (2HDM) and/or minimal supersymmetric standard model (MSSM) [78]. It is also shown in Ref. [21] that using an effective field theory approach, this branching ratio $B(B_s \to \tau^+\tau^-)$ can be as large as $15.0 \times 10^{-2}$ whereas an enhancement up to $5.0 \times 10^{-2}$ is possible in the model with an extremely light $Z'$ boson. Our estimation gives $B(B_s \to \tau^+\tau^-) = 1.64 \times 10^{-4}$ for a $Z'$ boson with mass = 100 GeV, which could help to explain the dimuon charge asymmetry observed in the B system. However, the contributions from NP in the $B_d$ sector are essential for full explanation [79,80].

## 5. Concluding Remarks

We have investigated the effect of both Z and $Z'$-mediated FCNCs on $B_s \to \tau^+\tau^-$ decay. Our predicted branching ratio value is enhanced relative to SM prediction and hence provides signals for new physics beyond the SM. Moreover, our predicted branching ratio value matches the recent LHCb collaboration limits. In this paper, we have studied the mass of $Z'$ boson in the range of 100 GeV – 2 TeV. We observe that as the mass of $Z'$ boson increases, the contribution of $Z'$ boson to the branching ratio decreases. For a light $Z'$ boson with mass = 100 GeV, we find an enhancement up to $B(B_s \to \tau^+\tau^-) = 1.64 \times 10^{-4}$ is allowed. This enhancement of $B(B_s \to \tau^+\tau^-)$ [20,21] could help to explain the recently observed like-sign dimuon charge asymmetry to some extent. In order to fully reconcile the D0 result, some contributions from NP in the $B_d$ sector are needed [79,80]. We hope the LHCb would provide much more precise data on it very soon.

## Acknowledgments

We would like to thank Diptimoy Ghosh, Tata Institute of Fundamental Research, India. and Prafulla Kumar Behera, Indian Institute of Technology, Madras, India for fruitful discussions and suggestions. We thank the reviewer for suggesting valuable improvements of our manuscript.

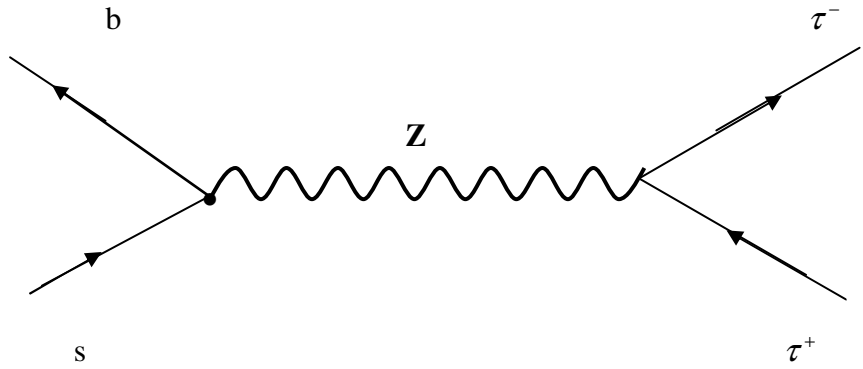

Fig.1 : Feynman diagram for $B_s \to \tau^+ \tau^-$ in a model with tree level FCNC transitions, where the blob ( • ) represents the tree level flavor-changing vertex.